# Automatic Detection and Classification of Corona Infection (COVID-19) from X-ray Images Using Convolution Neural Network

Kinjal A. Patel[1],
[1]Department of Computer Application and IT
[1]Gujarat Law Society University
[1]Ahmedabad, India

Tanvi Goswami[2],
[2]Department of Information Technology
[2] Dharmsinh Desai University
[2]Nadiad, India
[2]goswami.tanvi@gmail.com

*Abstract*— The novel coronavirus universally known as the COVID-19 outbreak arises at the end of 2019 in one of the East Asian countries and it is subjected to widespread discussion and debate. There are almost 200 countries affected across the globe by COVID-19. And, it has ruined many lives and the global economy. The virus is spreading very rapidly at the pace of around 10 fold in less than a month. Also, in the case of COVID-19, it is critical to detect the infection, as it employs various symptoms which may differ from person to person. Hence, diagnosis in starting stage and treatment are very much important for such type of infectious disease. The chest x-ray is one of the primary techniques among blood tests and Computed Tomography contributes a major role in the early diagnosis of COVID-19. There is a rising need for automated and auxiliary diagnostic tools for early diagnosis, as there are no accurate and truthful automated toolkits on hand. In this resarch study, we have designed a Convolution Neural Network architecture – a deep net for the classification of x-ray images of chest among two classes: COVID-19 or Non-COVID- 19 infection. The anticipated model is expected to provide accurate diagnostic results and produced classification accuracy of 99%, 100%, and 100% with 70%-30%,75%-25% and 80%-20% train-test data split respectively, for the binary classification of the x-ray image to be COVID-19 or Non-COVID-19 infection category. We have designed the CNN with optimized parameters with 3 convolution layers and optimized number of filters in each layer.

## I. Introduction

The coronaviruses (CoVs) are most infectious in birds but, from several decades it is capable to change host and infecting the humans as well. The coronavirus also known as SARS-CoV-2 is a new member of SARS (severe acute respiratory syndrome) virus spices recognized by its genome sequences [1]. The novel coronavirus epidemic arises in one of the East Asian countries during December 2019 and has extensively blowout across the countries. The coronavirus is rapidly spread from humans to humans. The novel coronavirus indications vary from one person to another person. The extreme symptoms of the COVID-19 are high body temperature, dry cough, and weariness as well as few symptoms are sore throat, diarrhea, and headache [2]. The serious symptoms of the COVID-19 are chest aching and struggle in breath [2]. There are 7818 coronavirus positive cases and 170 deaths are confirmed across the globe during the end of January [3]. The "Public Health Emergency of International Concern" has been declared by The World Health Organization (WHO) on Janyary 30, 2020 due to the extensive increment in coronavirus positive cases [4]. WHO entitled this disease as COVID-19 by February 2020 [5]. As of 25 May 2020, 53,04,772 COVID-19 positive cases and 3,42,029 deaths are confirmed across the globe and 1,50,762 COVID-19 positive cases, and 4,349 deaths are confirmed across India [6] [7].

The COVID-19 is transmissible, thus many countries are applied lockdown to prevent the spreading of COVID-19. Due to this kind of situation, the Government of India has launched many competitions to get innovative ideas from intelligent minds to deal with every domain such as patient care management, fake news detection, movement tracking, large scale sterilization, stabilizing affected business, virus containment and video conferencing for online education [8]-[11].

The numbers of researchers are working on finding proper treatment, antivirus drugs, and therapeutic vaccines of COVID-19. The detection of COVID-19 infection at a premature stage and isolation of the disease infected people is mandatory as the antivirus drug and therapeutic vaccine was not available. The unexpected hike in a positive COVID-19 leads to the need for a massive amount of medical resources. The real-time reverse transcription-polymerase chain reaction (RT-PCR) test is used to detect COVID-19 infection. But, RT-PCR test takes more time to give the result as well as the result is not precise. RT-PCR has high false-negative rates and low sensitivity [12] [16]. The less accurate RT-PCR is not adequate in the present epidemic condition. In many cases, the negative results are acquired from the RT-PCR test, and COVID-19 infected person may not be identified, thus the healthy people may get infected from COVID-19 positive people. The indication of COVID-19 may include dry cough and chest pain and it can be examined using radiological imaging of the lungs and chest, computed tomography (CT) and X-ray. The CT has high sensitivity and accurate results [12]. As the CT is more accurate therefore it is useful to detect the infection of COVID-19 at an early stage. The radiologists examine the CT and X-ray images. Due to the hike in COVID-19 positive cases and shortage of medical professionals as well as RT-PCR tool kits, there is a requirement for the development of an automatic tool that detects the COVID-19 infectivity from CT scans and X-ray imagery. The deep learning based method come up to classify the COVID-19 and Non-COVID-19 infections from CT and X-ray images is worth advantageous while a shortage of medical professionals.

The core objective of this paper is to classify COVID-19 and Non-COVID-19 diseased person from X-ray images of the chest. The deep learning approach is used to categorize infected COVID-19 patients and non-infected COVID-19. The 2 dimensional convolution neural network (CNN) is designed for binary classification which classifies the X-ray images into two classes namely; COVID-19 and Non COVID-19. The CNN model is implemented with three 2D - convolutional layers and trained by chest X-ray images. The model performance is evaluated with different evaluation measures such as precision, recall, f-score, accuracy, and confusion matrices.

The outline of the paper is as follows: the work has been completed in the domain of COVID-19 is discusses in "Literature review" section; the implemented CNN model for classification is discussed in the "Proposed CNN model" section; the performance of CNN model and its comparison with existing models is discussed in the "Performance comparison and analysis" section; the paper is accomplished in the "Conclusion" section.

## II. LIETRATURE REVIEW

The researchers have discovered the identity of the COVID-19 from patterns on images of chest CT and X-ray [16]-[24]. Due to the shortage and less sensitivity of RT-PCR kits, various comparative surveys are conducted with RT-PCR test and chest CT and X-ray. The study was conducted by Xie et al. [14] to identify COVID-19 infectious from 167 people. A total of 3% of 167 people had negative RT-PCR for COVID-19 infectious. But, that 3% of people were founded positive COVID-19 infectious on chest CT. The other study was conducted by Berheim et. al [15] by examining 121 chest CTs of COVID-19 infected patients. The increased severity of infection is founded in concerning time from initial infection was detected. Fang et al. [13] examined the 2 patients along with their symptoms as well as travel history. They came up with the results that CT of chest had a higher sensitivity as compared to RT-PCR to identify COVID-19.

Ozturk et al. [16] developed the DarkCovidNet model which has 17 convolutional layers with different filters at every layer. They used DarkCovidNet for 2 classes namely; for COVID and No-findings as well as the multi-class classification for Pneumonia, COVID, and No- findings. The model formed 98.08% of classification accuracy for 2-class classification and 87.02% of classification accuracy for multi-class classification. Singh et. al [17] developed a CNN to classify COVID-19 positive infection or not from CT images of chest. They tuned the parameters of CNN by using multi-objective differential evolution (MODE). The MODE based CNN gives a significant performance with 1.9789% accuracy, 2.0928% F-measure, 1.8262% sensitivity, 1.6827% specificity, and 1.9276% Kappa statistics.

Khan et al. [18] extract the features from x- ray images of the chest of COVID-19 patients and fed that features into the SVM for classification of positive COVID-19 or negative COVID-19. The developed COVID-RENet and Custom VGG model give accuracy 98.3%, AUC 0.98, F-score 0.98, Recall 96.67%, and Precision 100%. Wang et. al [19] proposed CNN as a COVID-Net which as lower complexity of architecture and computation as compare to VGG-19 and ResNet-50. The COVID-Net was already trained on the ImageNet and then again trained that model on the COVIDX dataset to categorize Pneumonia, COVID-19, and normal patients. The 93.3% classification accuracy was obtained from custom developed model COVID-net and on the other hand, 83.0% classification accuracy obtained from VGG-19 and 90.6% classification accuracy obtained from ResNet-50.

Narin et al. [20] have used ResNet50, InceptionV3, and Inception-ResNetV2 to categorize between normal X-ray images and COVID-19 infected images. To overcome the shortage of sufficient data, they used the ImageNet dataset as well. Also, transfer learning technique was employed which also reduces the time needed for training the model. The 98% of classification accuracy is obtained by pertained ResNet-50 model whereas 97% and 87% of classification accuracy obtained by InceptionV3 model and Inception- ResNetV2 respectively. Hasan et al [21] has projected a Q-deformed entropy algorithm based on deep learning for feature extraction from lung images. These extracted features fed into LSTM) neural network which classifies: Pneumonia, COVID-19 and healthy lungs. They obtained 99.68% accuracy from the projected model.

Apostolopoulos et al. [22] has used various existing pre-trained CNN architecture which are: MobileNet V2, VGG 19, Xception, Inception Resnet v2 and Inception. They come up with the comparative study of these models and applied transfer learning to classify viral and bacterial pneumonia, COVID-19, and also normal lungs conditions from X-ray images. There study includes two datasets. Dataset-1 with two classes' normal and COVID-19 and Dataset-2 with three classes normal conditions, viral pneumonia and COVID-19. The highest accuracy is obtained in VGG-19 which is 98.75% for dataset-1 and 93.48% for dataset-2 among all the models. The MobileNet V2 obtained 97.40% and 92.85% accuracy, Inception obtained 86.13% and 92.85%accuracy, Xception obtained 85.57% and 92.85% accuracy and Inception Resnet v2 obtained 84.38% and 92.85% accuracy for dataset-1 and dataset-2 in that order. Abbas et al. [23] uses custom-developed CNN – DeTrac known for Decompose, Transfer, and Compose to classify the COVID from normal cases and SARS cases by using X-ray images. The concept of DeTrac handles the irregular image dataset. They obtained accuracy of 95.12%, sensitivity 97.91%, specificity 91.87%, and precision 93.36%. Khalid et al. [24] has come up with comparative analysis of various CNN pre-trained architectures which are: DenseNet201, VGG16, Inception ResNet V2, VGG19, Resnet50, MobileNet V2 and InceptionV3 that detect and classify normal, COVID-19 and bacterial pneumonia chest X-ray images of chest and CT scan images. They obtained best accuracy of 92.18% from and 88.09% with Densnet201.

From this comprehensive analysis of existing studies, it is concluded that the CT images and X-ray images is likely to be considered for detecting and classifying the infection at its early stage [25]. Hence, in this paper, Convolutional Neural Network – deep learning model is implemented to classify COVID-19 patients from chest X-ray images.

III. PROPOSED WORK

A. Dataset Collection

Data is the primary step for developing any computer-aided diagnostic tool. The dataset for this experiment was gathered from [26]. The Dataset is having 912 images for each class. These images are original x-ray images as well as augmented images for two classes namely: COVID-19 and Non-COVID-19 detection from X-ray images of chest. COVID-19 images are original and augmented X-ray images whereas non-COVID-19 images are original images [26]. The sample image from each class is demonstrated in Fig. 1.

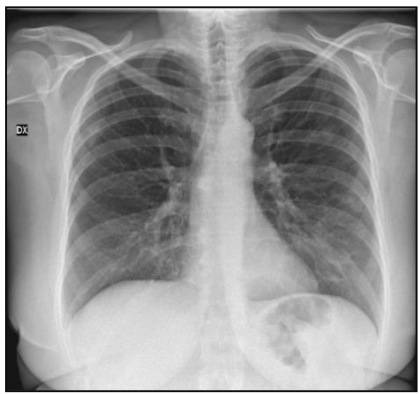
(a)

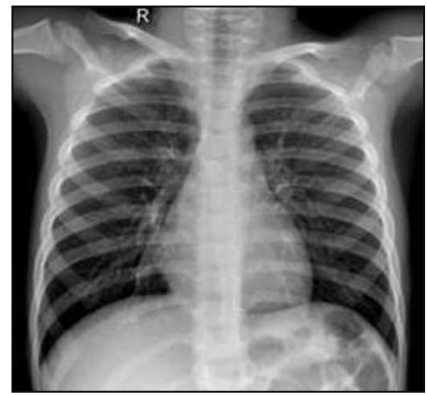
(b)

Fig. 1 (a) COVID-19 image, (b) Non-COVID-19 image [26]

*Proposed CNN Architecture*

We have observed and studied various parameters and hyper-parameters of CNN that is; Number of convolutional filters, number of convolutional layers, kernel size, pooling window size, fully connected layers, and number of neurons in fully connected layers. These parameters are selected after performing an analysis of works carried out by observing available pre-trained architectures used by researchers so far [19] [20]. As an outcome of the analysis, we have designed CNN architecture from scratch by keeping all the parameters of CNN into consideration.

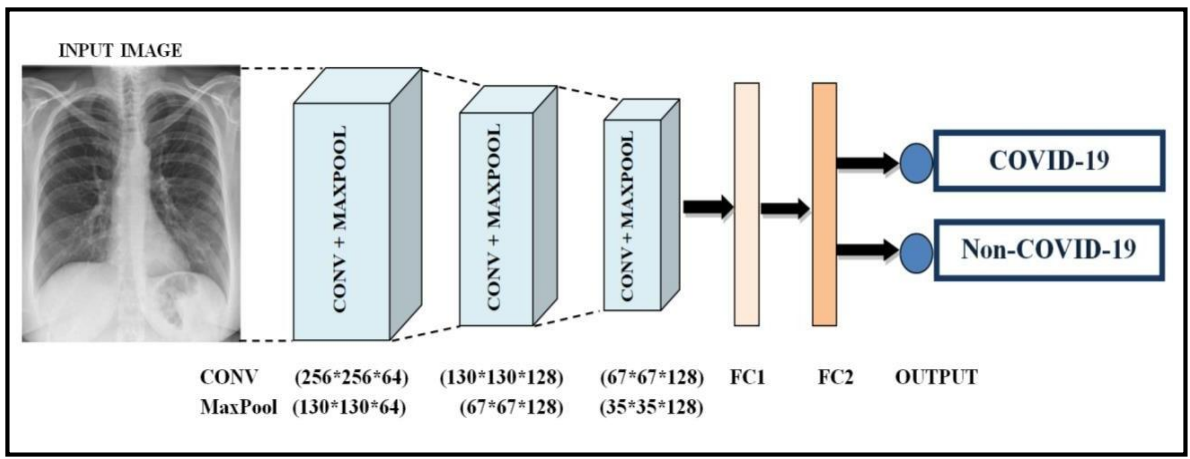

The proposed architecture of CNN is presented in Fig. 2. Our proposed model consists of three convolutional layers followed by 3 max-pooling layer, 2 fully connected layers, and an output layer consist of softmax activation fuction. We evaluated our model on the augmented COVID-19 X-ray image dataset [26]. We split the dataset into three train and test sets: 70% - 30%, 75% - 25% and 80% - 20% train-test split. The model was trained for 45 epochs. The default input image size of 256 × 256 was considered in the proposed model evaluation. Resizing is performed to change the various image dimensions to 256 × 256. The optimizer used was Adam 0.001learning rate throughout the experiments because Adam performs better than Gradient Descent in many machine learning applications. Drop out method was also used to avoid overfitting, which randomly drops connections from the previous layer. The values of Adam's learning rate and the number of epochs were selected after certain observations of training loss. The architecture details are listed in Table 1.

TABLE I
CONFIGURATIONS OF PROPOSED THE CNN ARCHITECTURE

| Layer | Processing unit | Input Dimension | Output Dimension | Kernel Size | No. of Kernels | Zero padding |
|---|---|---|---|---|---|---|
| 1 | Conv2D | 256x256 | 256x256x64 | 3x3 | 64 | 2x2 |
| 2 | MaxPool | 256x256x64 | 130x130x64 | 2x2 | 64 | - |
| 3 | Conv2D | 130x130x64 | 130x130x128 | 3x3 | 128 | 2x2 |
| 4 | MaxPool | 130x130x128 | 67x67x128 | 2x2 | 128 | - |
| 5 | Conv2D | 67x67x128 | 67x67x128 | 3x3 | 128 | 2x2 |
| 6 | MaxPool | 67x67x128 | 35x35x128 | 2x2 | 128 | - |
| 7 | Fully Connected Layer 1: 1024 neurons followed by activation ReLu and 0.2 DropOut rate | | | | | |
| 8 | Fully Connected Layer 2 :1024 neurons followed by activation ReLu and 0.2 DropOut rate | | | | | |
| 9 | Output Layer with softmax activation function | | | | | |

To avoid shrinkage of dimensions, zero-padding with size 2x2 is used. Steps per epoch are also one of the hyperparameters of CNN architecture; its value can be determined by batch size as well as the number of training samples. A large batch size provides a better gradient and prevents jumping around, but too large batch size may create memory problems. After analyzing the available memory we kept 64 batch sizes. Also, we kept the kernel size 3x3 across the experiments.

## IV. RESULTS AND DISCUSSION

### A. Performance Evaluation

The dataset was divided into three different splits; 70% - 30%, 75% - 25% and 80% - 20% train- test splits respectively. We have designed the CNN architecture as shown in Fig. 2. The model is trained for 45 epoch. The results are summarized in Table 2.

The confusion matrix are shown in Fig. 3 with COVID-19 and Non-COVID-19 test results of the designed architecture.

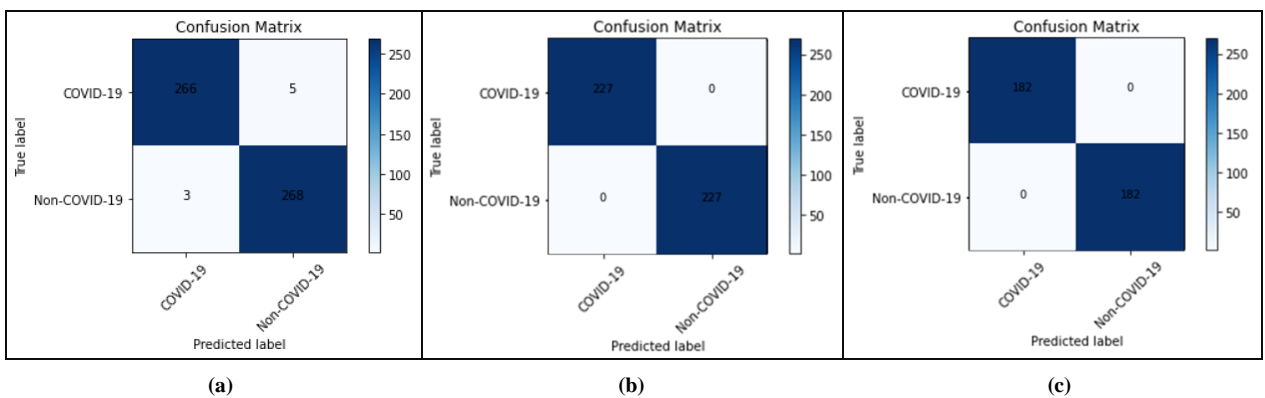

(a)      (b)      (c)

Fig. 3 The confusion matrix was obtained for three different splits using the proposed CNN model. The results for three experiments are shown in: a) confusion matrix for 70%-30% split b) confusion matrix for 75%-25% split, c) confusion matrix for 80%-20% split

We have obtained the best performance with 100% accuracy, 100% recall, and 100% precision for 75%-25% and 80%-20% splits. The lowest performance values have been obtained in 70%-30% splits with 99% accuracy, 99% recall and 99% precision. As a result experiment number 2 and 3 as mentioned in Table 2 provides better results as compared to experiment number 1.

TABLE II
PREDICTION PERFORMANCE RESULTS FROM THE CNN MODEL WITH DIFFERENT SPLITS

| Experiment Number | Train-test Split | No. of training images | No. of Testing Images | Confusion Matrix and Performance Results in % over the test set | | | |
|---|---|---|---|---|---|---|---|
| | | | | Precision | Recall | F1 Score | Accuracy |
| 1 | 70%-30% | 1282 | 542 | 99 | 99 | 99 | 99 |
| 2 | 75%-25% | 1370 | 454 | 100 | 100 | 100 | 100 |
| 3 | 80%-20% | 1460 | 364 | 100 | 100 | 100 | 100 |

*B. Discussion*

For this research work, the experiments were performed using Google Colaboratory with Tesla K80 GPU for Tensorflow and Keras library. The COVID-19 dataset is very limited also few research works have been reported as it is new to everyone so far. The major advantages of the proposed CNN model are: - 1) The model aims to classify chest X-ray images without using the traditional methods. The traditional technique uses different feature extraction and segmentation methods additionally while CNN extracts the features automatically. 2) The automated the approach is very much useful in assisting experts in decision making.

V. CONCLUSION

In this paper, a COVID-19 and Non-COVID-19 infection classification model is implemented to classify the infected humans from chest X-ray images. Primarily, the chest X-ray dataset of COVID-19 and Non-COVID-19 is decomposed into 70%- 30%, 75%-25%, and 80%-20% train-test data split. The data from each train-test split is applied separately to the training of the CNN model. The CNN model is trained with 45 epochs for each train- test data split. As a final point, the comparisons are drawn from each CNN model by considering diverse fractions of training- testing dataset. All-inclusive experimental results disclose that the proposed model outperforms in terms of accuracy, precision, recall, F1-score by 99% for each in 70%-30% split and 100% for each in the remaining two data splits. Hence, the proposed model is beneficial for instantaneous COVID-19 disease classification from chest X-ray images.